\documentclass[11pt]{article}

\usepackage{microtype}
\usepackage{graphicx}
\usepackage{booktabs} 
\usepackage{caption}
\usepackage{subcaption}
\usepackage[sort&compress]{natbib}
\usepackage{lineno}
\usepackage{xcolor}

\usepackage{hyperref}

\usepackage{chemformula}
\usepackage{amsfonts} 

\usepackage[margin=2.5cm]{geometry}

\usepackage{hologo}
\usepackage{mhchem}

\begin{document}

\title{Rapid modelling of reactive transport in porous media using machine learning: limitations and solutions}

\author{Vinicius L S Silva$^{*,a,b,c}$ \and Geraldine Regnier$^{b,f}$\and Pablo Salinas$^{b,d}$ \and Claire E Heaney$^{a,e}$ \and Matthew Jackson$^{b}$ \and Christopher C Pain$^{a,e}$}

\maketitle

\begin{center}
\small \noindent $^a$Applied Modelling \& Computation Group, Imperial College London, UK\\
\small \noindent $^b$Novel Reservoir Modelling and Simulation Group, Imperial College London, UK\\
\small \noindent $^c$Petroleo Brasileiro S.A. (Petrobras), Rio de Janeiro, Brazil \\
\small \noindent $^d$OpenGoSim, Leicester, UK \\
\small \noindent $^e$Centre for AI Physics Modelling, Imperial-X, Imperial College London, 
UK\\
\small \noindent $^f$Suez International, Tashkent, Uzbekistan\\ 
\small \noindent $^*$ Corresponding author email: v.santos-silva19@imperial.ac.uk
\end{center}

\begin{abstract}
Reactive transport in porous media plays a pivotal role in subsurface reservoir processes, influencing fluid properties and geochemical characteristics. However, coupling fluid flow and transport with geochemical reactions is computationally intensive, requiring geochemical calculations at each grid cell and each time step within a discretized simulation domain. Although recent advancements have integrated machine learning techniques as surrogates for geochemical simulations, ensuring computational efficiency and accuracy remains a challenge. This work investigates machine learning models as replacements for a geochemical module in a simulation of reactive transport in porous media. As a proof of concept, we test this approach on a well-documented cation exchange problem. While the surrogate models excel in isolated predictions, they fall short in rollout predictions over successive time steps. By introducing modifications, including physics-based constraints and tailored dataset generation strategies, we show that machine learning surrogates can achieve accurate rollout predictions. Our findings emphasize that even for a simple sorption equilibrium reaction (cation exchange problem), machine learning surrogates alone fail in predicting over successive time-steps. Incorporating simple physics-based modifications enables us to overcome this limitation. A detailed analysis of the limitations and potential mitigation strategies is presented in this work. 

\end{abstract}

\textbf{Article Highlights}
\begin{itemize}


    \item Various machine learning and deep learning strategies are evaluated to determine their suitability for geochemical computations.
    
    \item Machine learning surrogates excel in isolated predictions but require modifications to avoid failure in successive time steps.
    \item Simple and non-intrusive physics-based corrections are proposed to enhance machine learning surrogate accuracy in successive time-step predictions. 
    

    
    
\end{itemize}


\section{Introduction}

Reactive transport within porous media occurs in a range of subsurface processes including geothermal energy extraction, hydrocarbon recovery, CO$_2$ storage, and groundwater flow. These phenomena can significantly affect the fluid dynamics, as well as the petrophysical and geochemical properties of the reservoirs \citep{appelo:04geochemistry,fan2012fully,oliveira2019modelling,kala2020element,yekta:21reactive,soulaine:21porousmedia4foam,nogues2013permeability,molins2012investigation}. Nonetheless, simulating fluid flow and transport coupled with chemical reactions is highly computationally intensive, due to the necessity of computing chemical reactions within each discrete element of the simulation grid, which can range from hundreds to millions of elements in size. 


Two predominant methodologies exist for solving transport and reaction processes together: the sequential method \citep{oliveira2019modelling,yekta:21reactive,soulaine:21porousmedia4foam}  and the fully coupled (or simultaneous) method \citep{fan2012fully,kala2020element,seigneur2023compositional}. In the sequential method, transport and chemical reactions are solved in a stepwise manner. The fully coupled method, on the other hand, simultaneously resolves both transport and reaction equations, iterating until convergence is achieved. Both methodologies necessitate executing geochemical calculations on a per-grid-block basis, which remains the primary contributor to the computational time. Even worse for the fully coupled scheme, where the calculations need to be performed for each iteration within a time step (until convergence).

The use of machine learning to address complex engineering challenges has paved the way for advancements in reactive transport simulations.
The trend of using machine learning as a substitute for conventional geochemical simulators is on the rise \citep{jatnieks:16data,li:19accelerating,leal:20accelerating,guerillot:20geochemical,de:21dectree,demirer2023improving,sprocati2021integrating} although ensuring computational efficiency and prediction accuracy remains an open challenge. Mainly, when we consider the accumulation of error through time steps (rollout error), since the solution of one time iteration is the input for the next one. Thus, any small deviation from the correct solution can be amplified and generate unbounded errors during the simulation run. 
\citep{jatnieks:16data,leal:20accelerating,de:21dectree,demirer2023improving} considered the rollout error when evaluating the machine learning surrogate, but they used the full-physics coupled simulation to generate the dataset for training the machine learning model, something we decided to avoid since the goal is to replace the full-physics coupled simulation. \citet{guerillot:20geochemical} reported the rollout error but did not use the coupled simulation to generate the dataset; however, they mention that their approach is not mass conservative, and further investigation should be done to solve this issue.    
To the best of the authors' knowledge, a detailed evaluation of machine learning models, training set sampling procedures, and the incorporation of physical constraints is still lacking for building surrogates of geochemical simulators.    

In this work, we explore machine learning techniques to develop a rapid surrogate model for a geochemical simulator. 
The aim is to facilitate an efficient integration with flow and transport simulators while reducing computational overhead. Our efforts concentrate on modelling cation exchange reactions between an aqueous solution and a rock surface in a porous medium scenario, a topic well documented in the literature \citep{appelo:04geochemistry,parkhurs:13description,silva2022rapid,yekta:21reactive}. We select this basic sorption equilibrium reaction as a proof of concept to conduct a detailed analysis of the limitations and advantages of using a machine learning surrogate. We assess various machine learning and deep learning strategies to determine their suitability for geochemical computations. We also investigate how the error evolves when multiple time steps are considered (rollout), the effect of different dataset sampling procedures on the prediction error, and how prior information or physical knowledge of the problem can help improve prediction accuracy. Our findings suggest that with careful design and consideration of limitations, machine learning surrogates can effectively substitute the cation exchange calculation in a reactive transport in porous media. 

In summary, we make the following contributions: 
\begin{itemize}
    \item We analyse different machine learning and deep learning models for tabular data to work as a surrogate of the cation exchange problem. We assess the machine learning models in terms of accuracy and speed-up of the geochemical calculation. 
    \item We investigate different strategies of dataset generation in order to quantify their effect on the rollout/simulation error. We propose alternatives for using the full-physics coupled simulation to generate the dataset, given that the final goal is to avoid the expensive coupling between the transport and geochemical simulators.    
    \item We perform a thorough investigation of the one-shot and rollout prediction errors. We demonstrate that machine learning surrogates excel in isolated predictions but fail in successive time steps without appropriate modifications. 
    \item We propose simple and non-intrusive (no need to access the simulator's code) physics-based corrections that enable the use of the machine learning surrogate in the cation exchange problem.   
\end{itemize}

The remainder of this paper is structured as follows. In Section~\ref{sec:testcase}, we describe the geochemical simulator, flow and transport simulator, coupling procedure, the cation exchange problem, and the machine learning/deep learning surrogates. Following that, we detail the results for the one-shot and rollout predictions in Section~\ref{sec:result}. Finally, a discussion and concluding remarks are provided in Section~\ref{sec:disc} and \ref{sec:conc}, respectively.  

\section{Method}\label{sec:testcase}

In this section, we first describe the geochemical simulator used in this work (PHREEQC). Following that, we present the flow and transport simulator (IC-FERST) and the coupling procedure used to integrate PHREEQC into IC-FERST. Then we describe the cation exchange problem, and finally we present the machine learning/deep learning models used as surrogates.    

\subsection{Geochemical simulator}

PHREEQC \citep{parkhurst:99user} is a public-domain geochemical reaction package developed and maintained by the United States Geological Survey. PHREEQC simulates geochemical processes including ion exchangers, equilibrium between water and minerals, solid solutions, and gases. 
Specifically tailored for embedding reaction calculations into diverse flow and mass transport simulators, PHREEQCRM is structured as a C++ class. Its core function involves receiving component concentrations from the transport simulator's grid cells, executing geochemical reactions, and subsequently relaying the updated component concentrations back to the transport simulator. In this work, we use PHREEQCRM in the coupling with the flow and transport simulator (IC-FERST).

\subsection{Flow and transport simulator}

We present the equations for incompressible single-phase porous-media flow. The pressure and velocity in the porous space have a linear relationship given by Darcy’s law as
\begin{equation}\label{eq:darcy}
\mathbf{u} = \frac{\mathbf{K}}{\mu} (-\nabla p + \rho g \nabla z ),  
\end{equation}
where $\mathbf{u}$ is the Darcy velocity and $\mu$ is the viscosity. $\mathbf{K}$ is the permeability tensor, $\rho$ is the density, $p$ is the pressure, $\nabla z$ is the gravity direction, and $g$ is the gravitational acceleration.

The continuity equation is given by
\begin{equation}\label{eq:massbal}
\nabla\cdot \mathbf{u} = 0.
\end{equation}  

We also include the transport equation for each species as
\begin{equation}
\phi \frac{\partial C_\alpha}{\partial t} + \nabla \cdot (\mathbf{u} C_\alpha) = 0
\end{equation}
where $C_\alpha$ is the concentration of species $\alpha$, and $\phi$ is the porosity. We enforce the flow of species into the simulation domain through the boundary conditions. 

We use IC-FERST for the solution of the equations reported here \citep{jackson:15,salinas:17b,silva:21machine}. IC-FERST is a next-generation three-dimensional reservoir simulator that uses surface-based modelling to represent the different petrophysical properties and unstructured meshes to discretize the domain. In surface-based modelling, the petrophysical properties are defined using discrete values bounded by surfaces \citep{jacquemyn2019surface}. In this approach, geological representation and the mesh used to discretize the domain are independent. 
IC-FERST uses for the space discretization a double control volume finite element method (DCVFEM) \citep{salinas:17b}, which is an improvement from the control volume finite element method (CVFEM) \citep{jackson:15,durlofsky:93,gomes:17}. 
Time is discretized using a $\theta$-method where $\theta$ varies between 0.5 (Crank-Nicholson) and 1 (implicit order) based on the total variational diminishing (TVD) criterion \citep{pavlidis:14}. 

\subsection{Coupling procedure}

\begin{figure}[!tb]
	\centering
	\includegraphics[width=\textwidth]{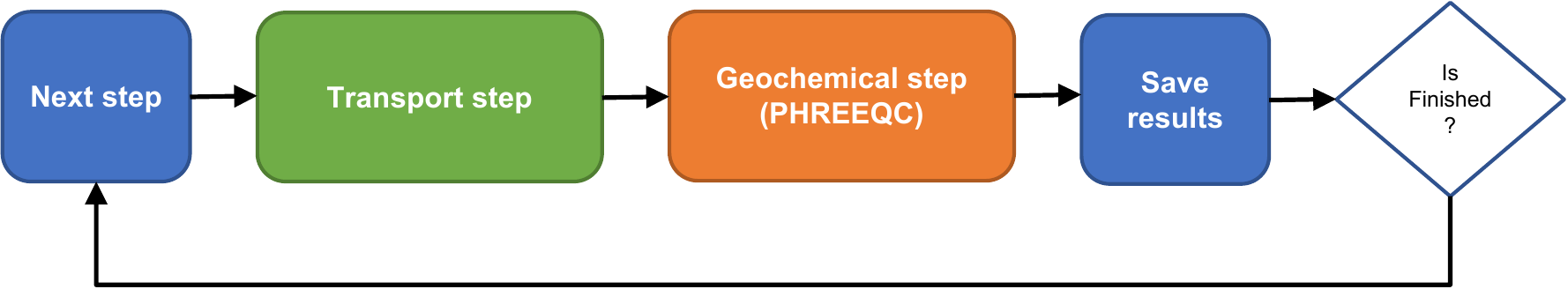}
	\caption{Coupling between the flow and transport simulator (IC-FERST) and the geochemical simulator (PHREEQC).}
	\label{fig:fc}
\end{figure}

We use a sequential approach to couple PHREEQC and IC-FERST in a non-iterative framework, see Figure~\ref{fig:fc}. The sequential non-iterative approach is a commonly used coupling scheme in reactive transport modelling \citep{oliveira2019modelling,yekta:21reactive,soulaine:21porousmedia4foam}. 
Here, the geochemical reactions (cation exchange) do not affect the characteristics of the rock or the flow behavior of the fluid. For other scenarios where the reactions can alter the porosity and permeability in the porous space or the viscosity of the fluids for example, a fully coupled scheme would be required \citep{fan2012fully,kala2020element,seigneur2023compositional}. 

In the coupling, IC-FERST controls the fields associated with the flow and transport, this includes pressure, velocity, saturation, and concentrations. Geochemical reactions are addressed by PHREEQC. Specifically, PHREEQC retrieves component concentrations from each grid cell within IC-FERST, processes the geochemical reactions, and then returns the updated component concentrations back to IC-FERST. 
The coupling is performed as follows:
\begin{enumerate}
    \item Configure the simulation domain and the related transport and chemical parameters initializing IC-FERST and PHREEQC. Set the boundary and initial conditions.
    \item IC-FERST updates pressure, velocity, and concentration (transport step in Figure~\ref{fig:fc}).
    \item PHREEQC computes the new concentration values for each grid cell (geochemical step in Figure~\ref{fig:fc}).
    \item  Advance to the next time level and repeat the procedure until the final time level is achieved.
\end{enumerate}

\subsection{Cation exchange problem}

Cation exchange reactions are sorption processes where the composition of water changes 
because of its contact with the rock surface. 
These surface reactions typically involve cations because the solid surfaces or substrates where the reactions occur are mostly covered with negative charges. The cation exchange is the modification of the cation composition of water through an exchange of cations between the solution and a solid surface. For example, a cation like \ce{Ca^2+} will attach electrostatically to the porous surface, displacing other cations such as \ce{Na+} into the solution. The cation exchange reactions are stoichiometrically balanced and can be represented as 
\begin{equation}\label{eq:cex}
    \ce{A\text{-}X + B+ <=> B\text{-}X + A+} \\ 
\end{equation}
where $\ce{A+}$ and $\ce{B+}$ are the cations, and $\ce{X}$ the rock surface (exchanger). The distribution of species is given by the law of mass action
\begin{equation}
    \ce{K_{B \backslash A} = \frac{[B\text{-}X][A+]}{[A\text{-}X][B+]}}.
\end{equation}
with [.] denoting activities (effective concentrations) \citep{appelo:04geochemistry}. $\ce{K_{B \backslash A}}$ is the equilibrium constant which defines the distribution at equilibrium of the species at the left and right side of Eq.~\ref{eq:cex}. If $\ce{K_{B \backslash A}}>1$ this implies that B is preferred over A (in the exchanger) and if $\ce{K_{B \backslash A}}<1$ then A is preferred over B. In the cation exchange, anions are not considered in the reaction equations since they do not interact with the exchanger. However, they can be transported through the domain in a reactive transport setup. Further details about the cation exchange formulation and solution can be found in \citet{appelo:04geochemistry} and \citet{parkhurs:13description}.   

\begin{figure*}[!tb]
	\centering
	\includegraphics[width=0.8\textwidth]{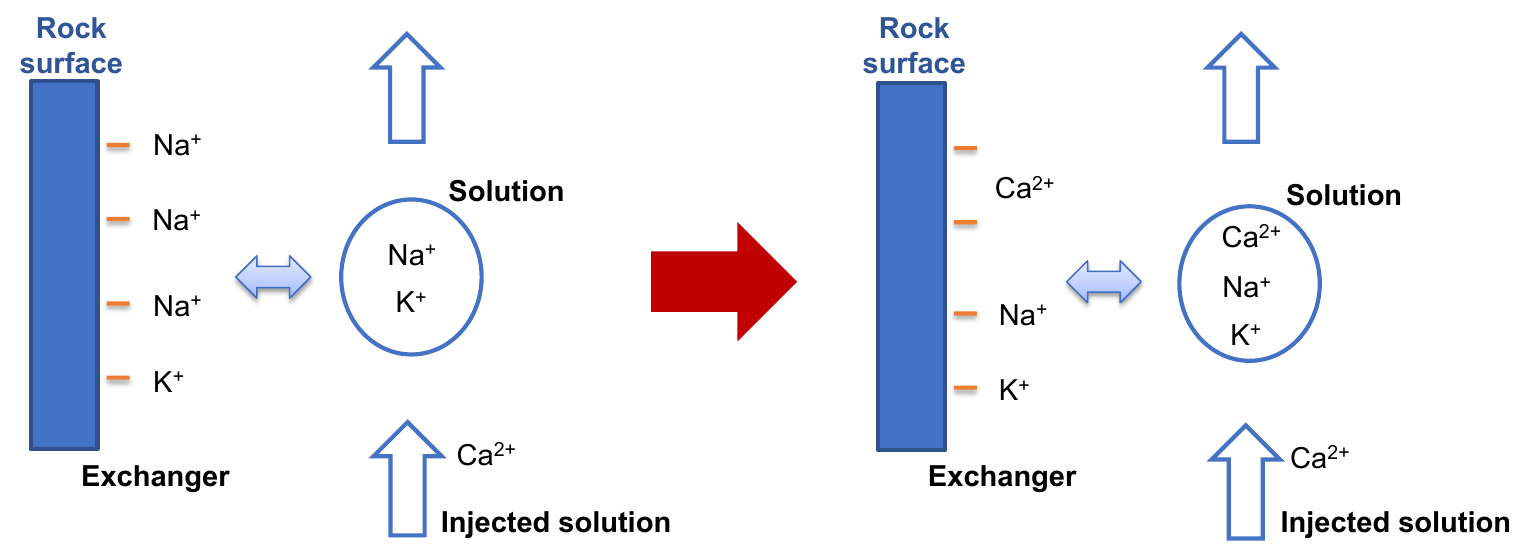}
	\caption{Schematic of the cation exchange in a small portion of the porous space. The outflow from the schematic will be the inflow in the next portion of the domain in the next time step.}
	\label{fig:ce}
\end{figure*}

\begin{figure*}[!tb]
	\centering
	\includegraphics[width=0.6\textwidth]{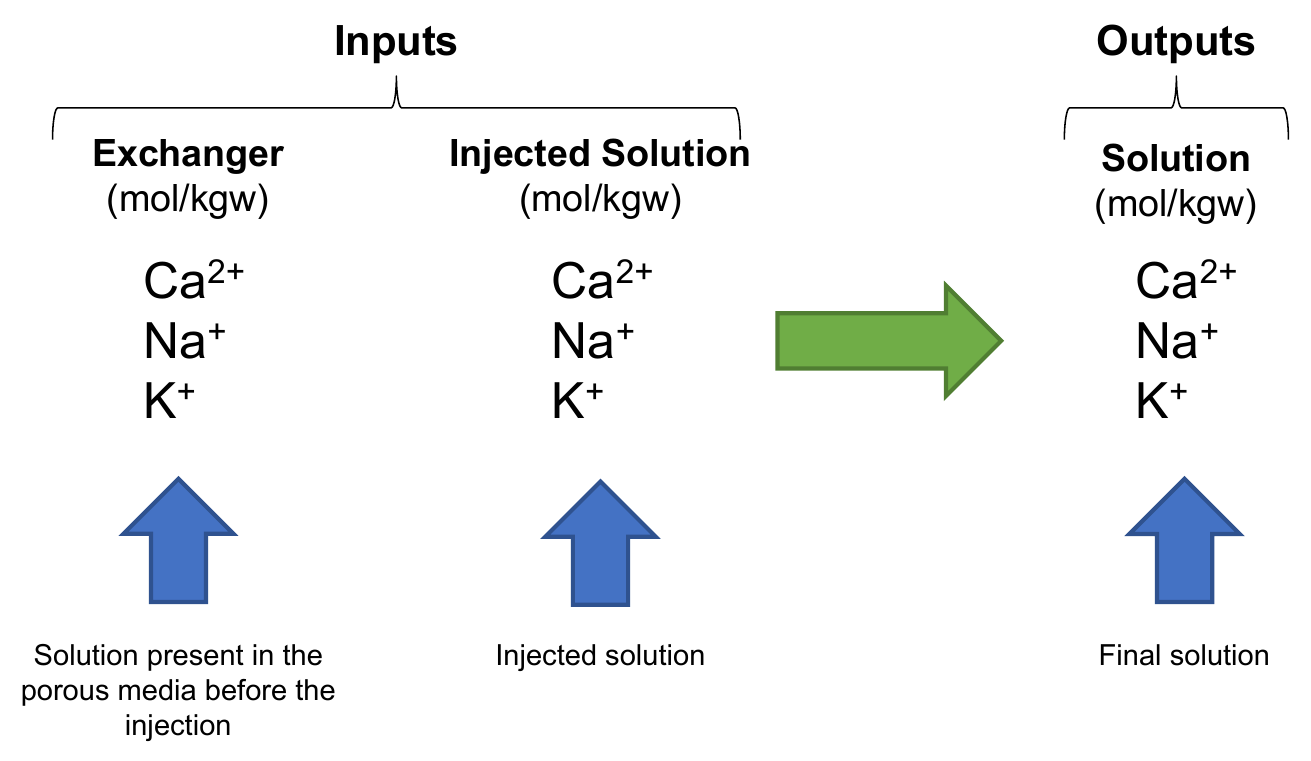}
	\caption{Inputs and outputs of the geochemical reaction in the cation exchange problem. The reaction is performed for each grid cell at each time step.}
	\label{fig:inout}
\end{figure*}

\begin{figure*}[!tb]
	\centering
	\includegraphics[width=1.0\linewidth]{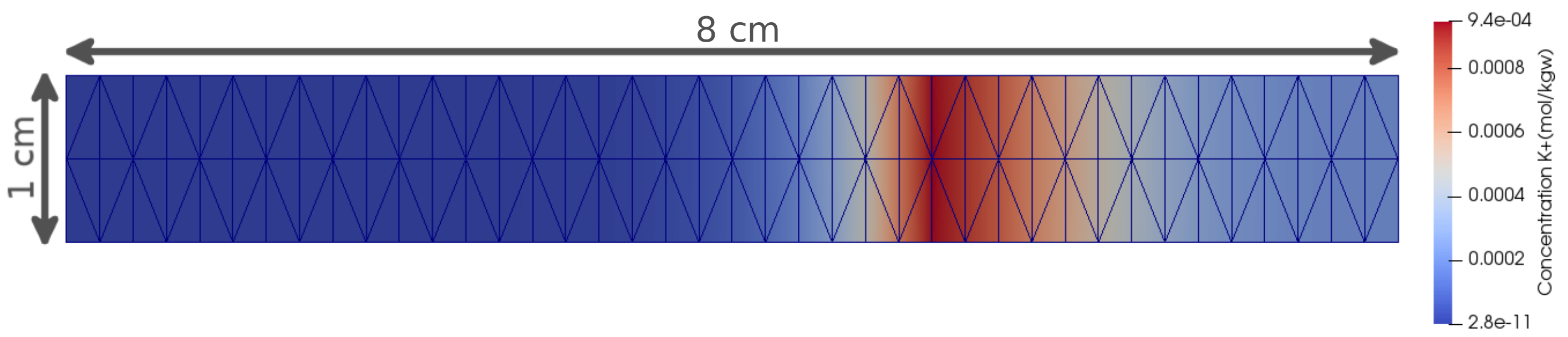}
	\caption{Two-dimensional domain and mesh used to represent the cation exchange problem (as in \citet{yekta:21reactive}). We show a snapshot in time of the concentration of \ce{K+}.}
	\label{fig:domain}
\end{figure*}

Figure~\ref{fig:ce} shows a schematic of the cation exchange problem considered here. The inputs of interest for the geochemical reaction are the concentration of the elements \ce{Ca^2+}, \ce{Na+}, \ce{K+} in the inflow solution and in the exchanger. The outputs are the new concentrations in the equilibrated aqueous solution, as shown in Figure~\ref{fig:inout}.  
To simulate this process, we use a two-dimensional domain shown in Figure~\ref{fig:domain}, alongside the mesh. We use the same model set up as in \citet{yekta:21reactive}. A calcium-chloride (\ce{CaCl2}) solution is flushed into the porous space at a constant rate of $2.78\times10^{-7} \text{m/s}$ from the left boundary, while the resulted solution is produced from the right boundary. The other boundaries are closed to flow. Table~\ref{tab:sol} shows the ion concentrations in the initial and injected solutions.

\begin{table}[!htb]
\caption{Initial and injected solution concentrations. The concentrations are reported in milimoles per kilogram of water (mmol/kgw).  }
\label{tab:sol}
\vskip 0.15in
\begin{center}
\begin{tabular}{lccccc}
\toprule
Solution &  \ce{Na+}  & \ce{K+} & \ce{Ca^2+} &  \ce{NO3-}  & \ce{Cl-}\\
\midrule
Initial & 1.0 & 0.2 & 0.0 &  1.2 & 0.0 \\
Injected & 0.0 & 0.0 & 0.6 & 0.0 & 1.2 \\
\bottomrule
\end{tabular}
\end{center}
\vskip -0.1in
\end{table}

\begin{figure}[!tb]
	\centering
	\includegraphics[width=0.9\linewidth]{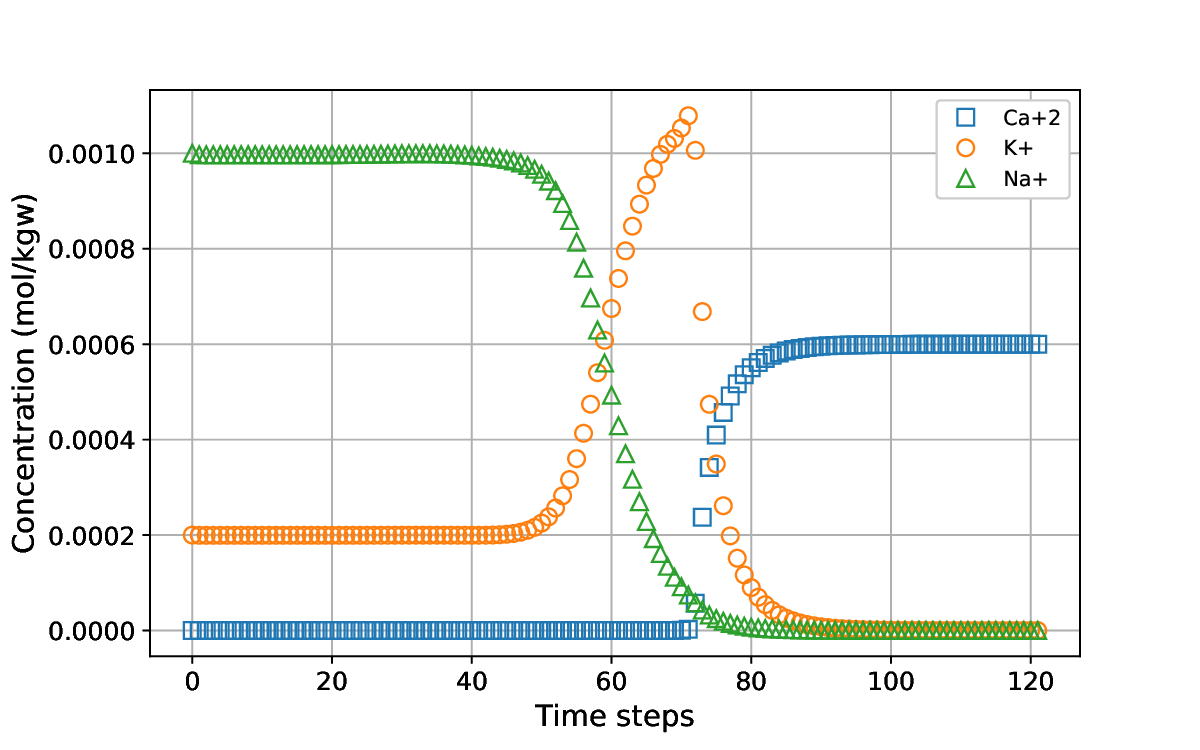}
	\caption{Cation exchange results generated by the coupling between IC-FERST and PHREEQC.}
	\label{fig:orig}
\end{figure}

In this scenario, as \ce{CaCl2} is introduced into the porous space, calcium (\ce{Ca^2+}) replaces sodium (\ce{Na+}) and potassium (\ce{K+}) from the solution and exchanger. As long as \ce{Na+} remains on the exchanger, the introduced \ce{Ca^2+} will continue to elute it. When \ce{Na+} is depleted, \ce{K+} is released from the exchanger, with its dissolved concentration rising to counterbalance the introduced \ce{Cl-}. Eventually, the \ce{Ca^2+} concentration stabilizes, matching the concentration of the incoming solution once all \ce{K+} has been discharged.
Figure~\ref{fig:orig} shows the outflow (right boundary) of \ce{Ca^2+}, \ce{Na+}, and \ce{K+}, over the reactive transport simulation, generated by the coupling between IC-FERST and PHREEQC, as in  \citet{yekta:21reactive}. Because there is no exchange of anions, they are only transported within the domain, we do not show them here.  
 
\subsection{Machine learning surrogates}\label{sec:ml}

\begin{figure*}[!tb]
	\centering
	\includegraphics[width=\textwidth]{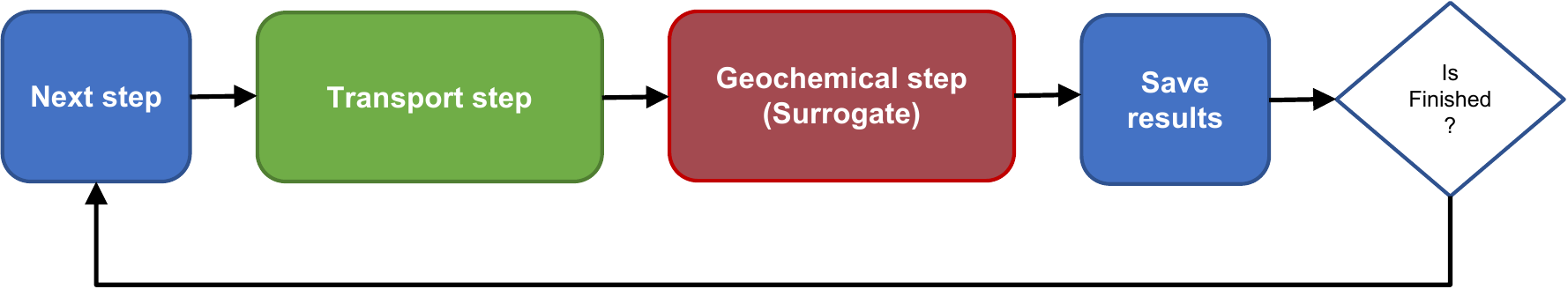}
	\caption{Coupling between the flow and transport simulator (IC-FERST) and the machine learning surrogate.}
	\label{fig:fcsurr}
\end{figure*}

In this work, we replace the geochemical simulator in the reactive transport by a machine learning surrogate, as shown in Figure~\ref{fig:fcsurr}.  The reaction calculations usually take the majority of the computational cost of the coupling between the numerical solution of flow and transport and the chemical reactions (up to 99\% in some cases) \citep{yekta:21reactive,li:19accelerating,jatnieks:16data,leal:20accelerating,de:21dectree,demirer2023improving,sprocati2021integrating}. The reaction calculations need to be performed for each grid cell of the discretized domain and for each iteration of the coupling procedure. Here, since we use a sequential non-iterative approach and consider that reaction equilibrium is reached within a time step, the chemical reactions are performed once at each time step (and for all grid cells).

We evaluate several machine learning models to identify the most suitable approach for the cation exchange problem. Although deep learning has seen significant advancements in recent decades, "classical" machine learning methods continue to outperform neural networks and deep learning models on tabular data \citep{shwartz2022tabular,borisov2022deep,grinsztajn2022tree}. Tabular data refers to data structured in rows and columns, such as the dataset used in this study, which is not in the form of images, text, or audio.

For this reason, we focus on testing "classical" machine learning methods, including linear regression, decision trees, random forests, and gradient boosting. Recent advancements in deep learning have introduced architectures based on the attention mechanism \citep{vaswani2017attention}, such as TabNet \citep{arik2021TabNet} and FT-Transformer \citep{gorishniy2021revisiting}, which have shown promising performance on tabular data. To ensure a comprehensive comparison, we include these attention-based models along with a standard neural network, the multilayer perceptron. A brief description of each method is provided below.

\textbf{Linear regression}: 
Linear regression predicts an output by calculating a weighted sum of the input features and adding a constant term \citep{hastie:09,geron:19}. The model produces a linear function of the input features. In this study, we use the linear regression implementation from \citet{scikit-learn}.

\textbf{Decision tree}: 
Decision trees use a hierarchical tree structure composed of nodes and branches to make predictions. They partition the feature space into a series of rectangular regions and assign a constant value to each region. Despite their simplicity, decision trees are powerful models and serve as the foundation for more advanced algorithms like random forests and gradient boosting \citep{hastie:09,geron:19}. In this work, we use the decision tree implementation from \citet{scikit-learn}.

\textbf{Random forests}: 
Random forests, introduced by \citet{breiman:01}, are ensembles of decision trees. When constructing the tree, each node's optimal split is determined from a randomly selected subset of features, and each tree is trained on a different sample of the dataset. These randomization steps help reduce the model's variance \citep{breiman:98, breiman:01}. 
The final prediction is obtained by averaging the outputs of all trees. In this study, we use the random forest implementation from \citet{scikit-learn}.

\textbf{XGBoost}: 
eXtreme Gradient Boosting (XGBoost) is an extension of gradient-boosted decision trees (GBDT) \citep{XGBoost}. GBDT builds an ensemble of decision trees sequentially, with each tree correcting the errors of the previous ones. XGBoost improves upon GBDT by introducing enhancements such as regularization, resulting in strong performance across many datasets \citep{shwartz2022tabular,borisov2022deep,grinsztajn2022tree}. It is one of the most widely used machine learning models for tabular data. The XGBoost library \citep{XGBoost} also includes a random forest implementation. In this study, we evaluate both the XGBoost and XGBoost random forest models.

\textbf{Multilayer perceptron}: 
A multilayer perceptron (MLP) is a fully connected feed-forward neural network composed of neurons with nonlinear activation functions. It includes at least three layers: an input layer, one or more hidden layers, and an output layer \citep{goodfellow:16,hastie:09}. In this work, the MLP is implemented using \citet{tensorflow:2015}.

\textbf{Tabnet}: 
TabNet is a deep learning model designed for tabular data, introduced by \citet{arik2021TabNet}. It employs a sequential attention mechanism to select which features to focus on at each step of its architecture. This feature selection is instance-specific, meaning it varies for each input. In this work, TabNet is implemented using \citet{paszke2019pytorch}.

\textbf{FT-transformer}: 
The FT-Transformer is an adaptation of the transformer architecture \citep{vaswani2017attention} for tabular data, introduced by \citet{gorishniy2021revisiting}. It employs a parallel attention mechanism from the transformer architecture and uses embedding layers for both categorical and numerical features. All features are transformed into embeddings, which are then processed through a stack of transformer layers. In this study, the FT-Transformer is implemented using \citet{paszke2019pytorch}.

\subsubsection{Residual connection}

Residual neural networks have achieved great success in computer video \citep{he2016deep} and recently in natural language processing (used in the transformer architecture) \citep{vaswani2017attention}. The main idea is to create a direct path from the input to the output (skip/residual connection). The layers in the machine learning model learn a residual function with respect to the input. Then we add the output of the layers to their input to generate the final results. This facilitates the training process and gradient propagation in neural networks \citep{he2016deep}. Here, we adapt the residual connections to the cation exchange problem by training some machine learning models to generate the change in concentration (delta) between the output and the injected solution (Figure~\ref{fig:inout}). Then to generate a prediction, we add the delta (output of the trained machine learning model) to the concentrations in the input (injected solution).  

\subsection{Hardware}

We run all experiments in a workstation with two AMD EPYC 7452 processors (32 cores each), 256 GB RAM memory, NVIDIA QUADRO RTX 4000 video card. 
During the training and inference of the machine learning models, we use all the cores available on the CPU or GPU (when supported). 
We report the prediction times for all machine learning models using only the CPU. This is because the machine learning surrogate needs to be integrated into the flow and transport simulator, that runs only on a CPU. 

\section{Results}\label{sec:result}



\subsection{One-shot prediction}

In this section, we evaluate the performance of the machine learning models described in Section~\ref{sec:ml}, considering a one-shot prediction. 
This means that we are not considering, in this section, the accumulation of error through time steps.  

\subsubsection{Machine learning model}\label{sec:mlm}

We compare several machine learning models (linear regression, decision tree, random forest, XGBoost random forest, XGBoost gradient boosting, multilayer perceptron, TabNet, and FT-transformer) to find the most suitable surrogate for the geochemical simulator. 
We run a hyperparameter optimization using a grid search with a 3-fold cross-validation for all machine learning models. We normalise all inputs using a min-max scaling between $-1$ and $1$. 
The mean square error (MSE) is used as a loss function in all cases and is given by
\begin{equation}\label{eq:mse_ml}
    MSE = \frac{1}{N_\alpha}\sum_{\alpha=1}^{N_\alpha} (C_\alpha - \Tilde{C}_\alpha)^2,
\end{equation}
where $C_\alpha$ is the concentration of species $\alpha$ in the output of the geochemical simulator, and $\Tilde{C}_\alpha$ the concentration in the output of the machine learning model. $\alpha$ represents the species $\{\ce{Ca^2+},\ce{Na+},\ce{K+}\}$, with $N_\alpha = 3$. We also define the root mean square error (RMSE) as 
\begin{equation}\label{eq:rmse_ml}
    RMSE = \sqrt{MSE}
\end{equation}

\begin{figure}[!tb]
    \centering
    \includegraphics[width=\textwidth]{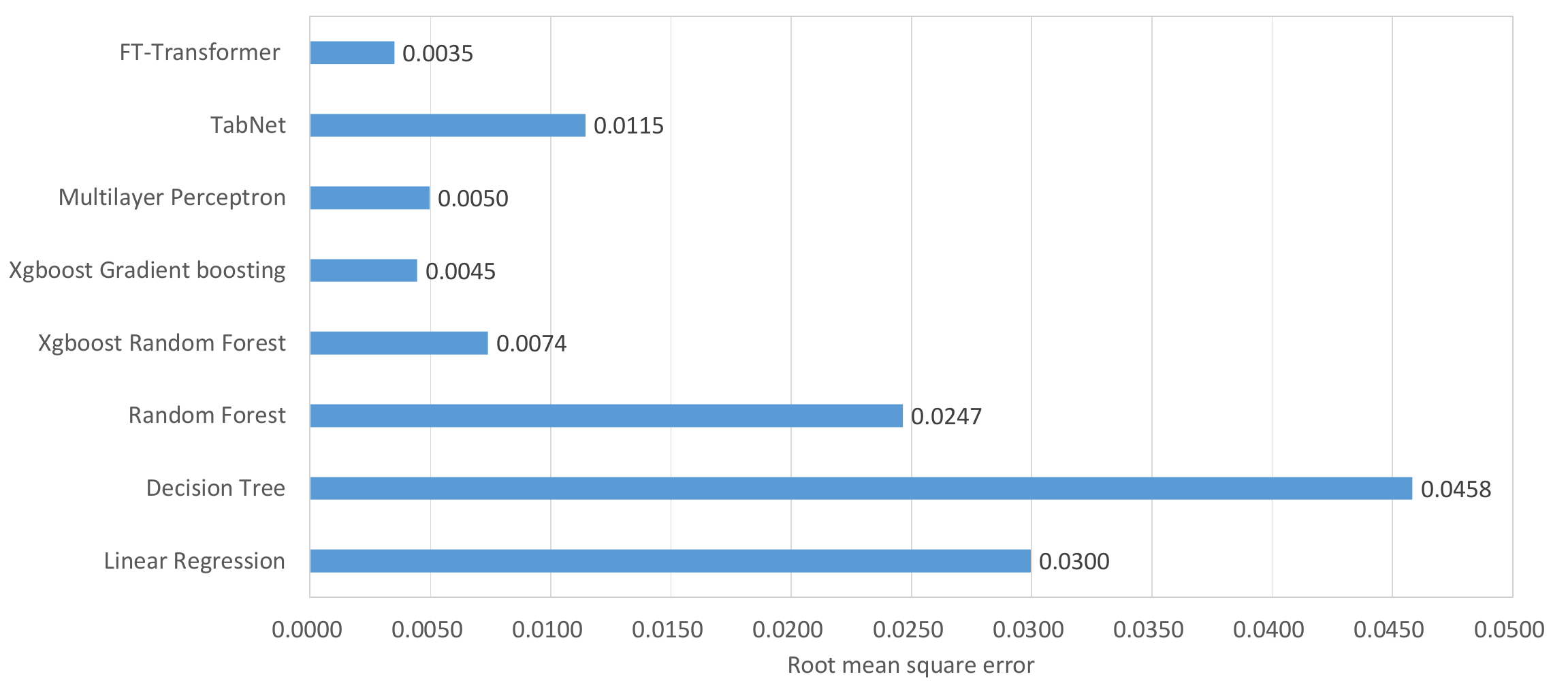}
    \caption{Root mean square error for different machine learning models.}
    \label{fig:mse1}
\end{figure}

\begin{figure}[!tb]
    \centering
    \includegraphics[width=\textwidth]{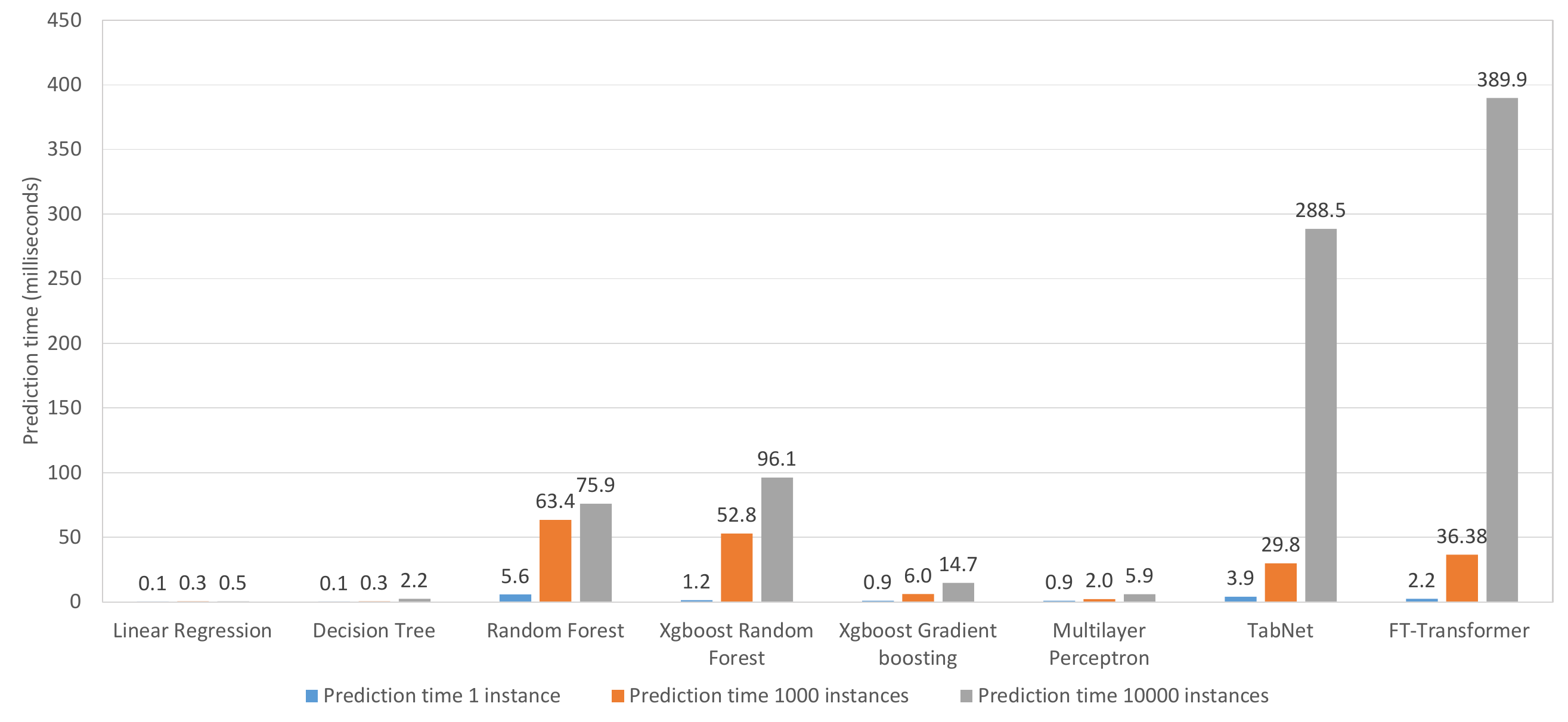}
    \caption{Prediction time for different machine learning models.}
    \label{fig:time1}
\end{figure}

\begin{figure}[!tb]
	\centering
	\includegraphics[width=0.7\linewidth]{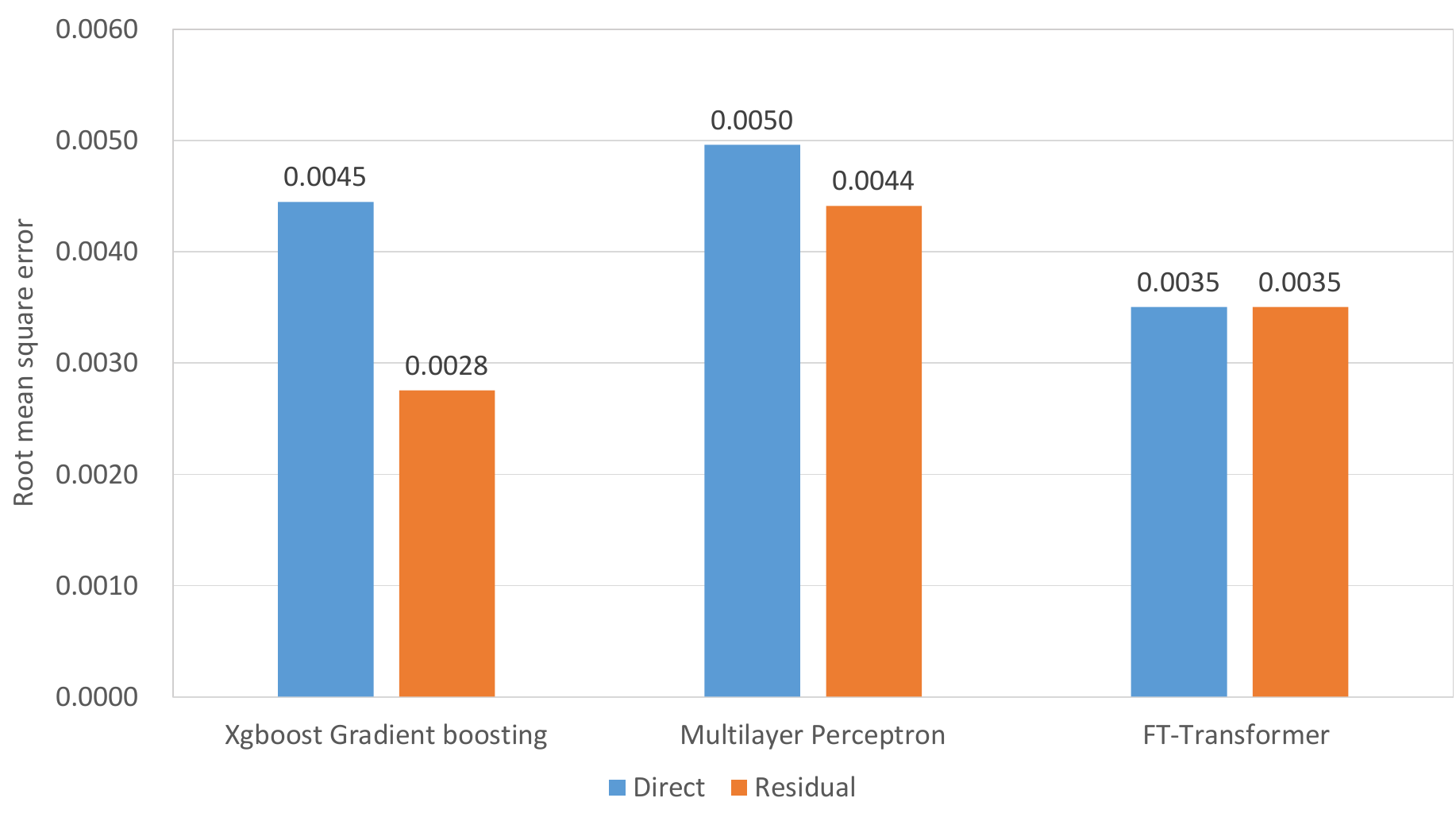}
	\caption{Comparison between the machine learning models with and without a residual connection.}
	\label{fig:res}
\end{figure} 

We start using a dataset that comprises $100,000$ instances generated by running the geochemical simulator with each input sampled from a uniform distribution, as in \cite{silva2022rapid}. We use $80\%$ of the dataset instances for the training and $20\%$ for testing. 
Figure~\ref{fig:mse1} shows the RMSE between the machine learning models and the geochemical simulator for the cation exchange problem. The minimum RMSE is from the FT-transformer, but XGBoost gradient boosting and the multilayer perceptron achieve a similar value. Figure~\ref{fig:time1} presents the prediction (inference) time of the models for different numbers of instances. The prediction was performed by calling the corresponding function in each model passing one two-dimensional array with the specified number of instances. For each model and number of instances, we run the prediction $1,000$ times and average the resulting prediction time. We can notice that linear regression, decision tree, XGBoost gradient boosting, and multilayer perceptron generate the best prediction times. The prediction time of one instance using the geochemical simulator (PHREEQC) is roughly 10 milliseconds. Considering that the goal is to replace the geochemical simulator in order to speed-up the reactive transport simulation, all machine learning models accomplish this goal since their prediction times are lower than the one from PHREEQC.  

The XGBoost gradient boosting and multilayer perceptron achieved the best compromise between prediction error and inference time. They can predict at least 10 times faster than the geochemical simulator. The FT-transformer has produced the lowest root mean square error, although a higher prediction time. In the next section, we will perform further investigations using these three machine learning models.

\subsubsection{Residual connection}  

\begin{figure}[!tb]
	\centering
	\includegraphics[width=1.0\linewidth]{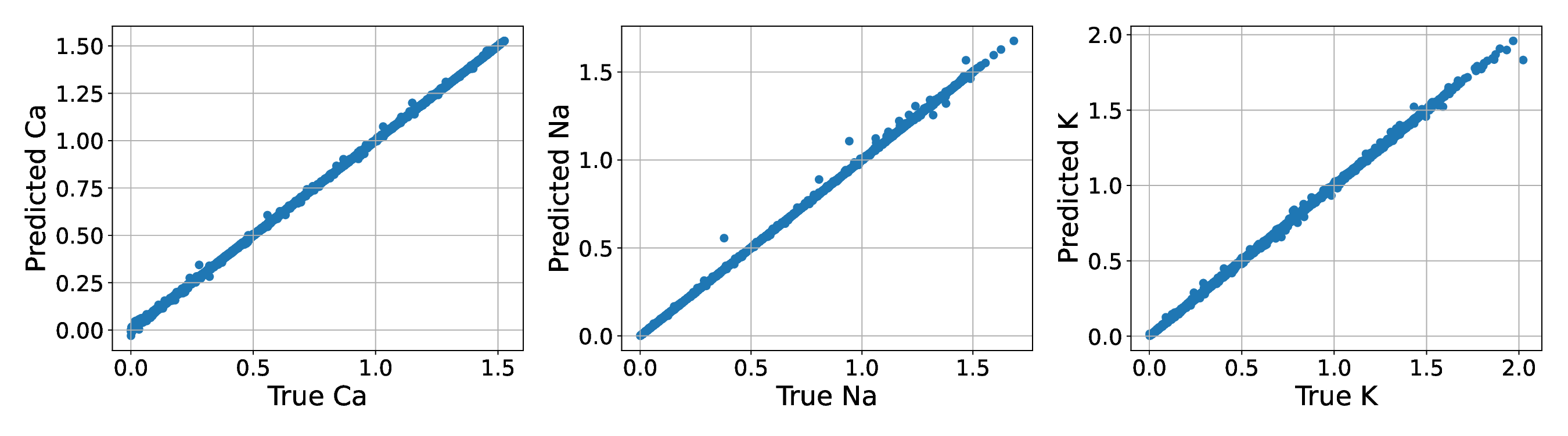}
	\caption{Prediction versus ground truth using the XGBoost gradient boosting with residual connection. Each plot shows the 20,000 points in the test set. The coefficient of determination is $R^2=0.993$.}
	\label{fig:pred}
\end{figure}

Figure~\ref{fig:res} shows a comparison between the three machine learning models selected in the previous section with and without a residual connection. The XGBoost and multlayer perceptron have reduced their prediction error by using the skip connection, while the FT-transformer produced the same result. The FT-transformer already has residual connections in its original architecture \citep{gorishniy2021revisiting} which explains the same error value. Considering the prediction error and inference time, we selected the XGBoost gradient boosting with residual connection to continue the experiments. 
Figure~\ref{fig:pred} shows the ground truth (geochemical simulator) versus the XGBoost predictions for the test set. The machine learning surrogate seems to accurately predict the cation concentrations.

\subsection{Rollout prediction}

In the previous section, we generated a machine learning surrogate to replace a geochemical simulator but only considered a one-shot prediction. However, during the numerical simulation the results from the geochemical reactions from one time step are used as inputs for the reactions in the next time iteration, and so on and so forth.  We call this a rollout prediction. The difference between the ground truth and the surrogate over the time steps, we call rollout/simulation error. It is calculated as the average of the RMSE (Eq.~\ref{eq:rmse_ml}) over all time steps. The rollout error appears because a prediction from one time step is used as a starting point for the prediction in the next time step. Hence, a small error or perturbation that occurs at the beginning of the simulation, can easily result in an unbounded rollout error.

\begin{figure}[!tb]
	\centering
	\includegraphics[width=0.9\linewidth]{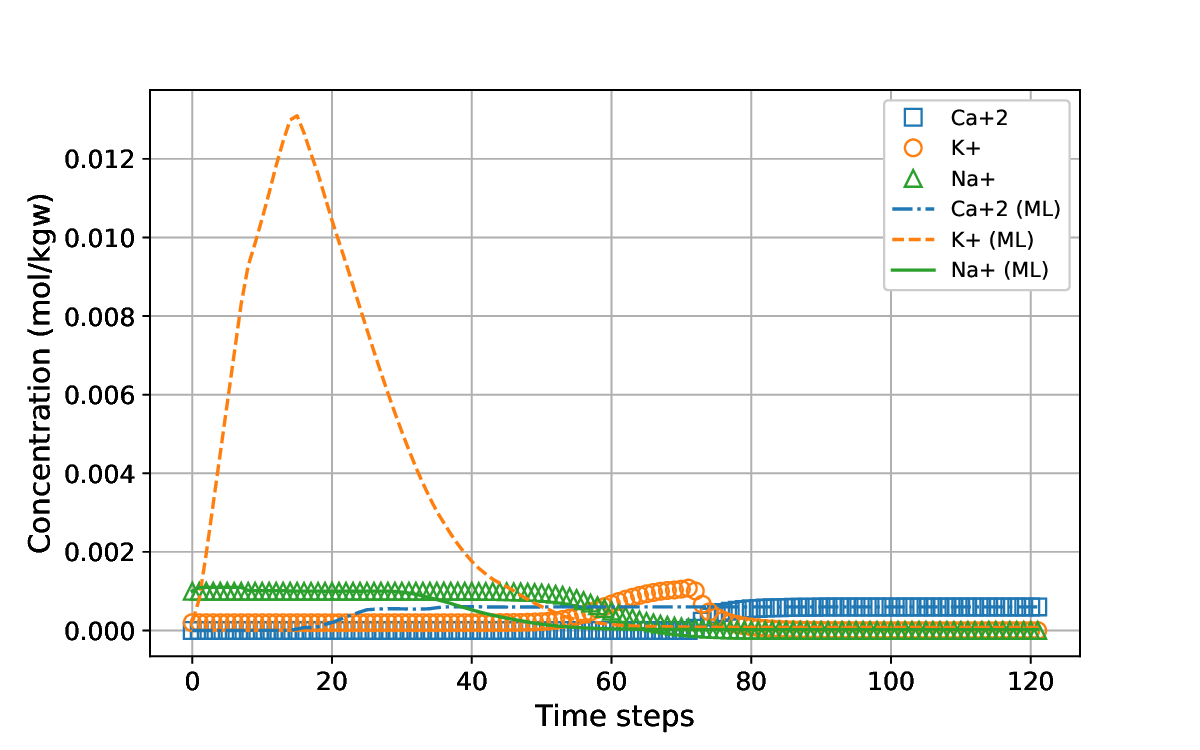}
	\caption{Simulation results comparing the coupling between PHREEQC (circles) and the machine learning surrogate (lines) with IC-FERST. Although the quality of the fit for the one shot prediction is good (Figure~\ref{fig:pred}), during the rollout the concentration of \ce{K+} generated by the machine learning surrogate explodes.} 
	\label{fig:pred1}
\end{figure}

We show in Figure~\ref{fig:pred1} the simulation (rollout) comparing the results from the flow and transport simulator (IC-FERST) coupled with the machine learning surrogate and PHREEQC.  We notice that even though the machine learning surrogate (XGBoost with residual connection) performed well for the one-shot prediction ($R^2=0.993$ in Figure~\ref{fig:pred}), when we consider the rollout, the simulation error escalates (the results are similar or worse when considering other machine learning models). Given these results, we decided to perform a thorough investigation of the rollout prediction using the machine learning surrogate.

\subsubsection{Dataset generation}

In the works from \citet{jatnieks:16data,demirer2023improving}, they mention the importance of the sampling procedure used to generate the dataset for the accuracy of the results in the coupling between transport and reactions. However, they do not give more details about this statement. Here, we will investigate different ways to generate the dataset, and the effect of the dataset size on the rollout/simulation error. One design choice was to not use the coupled simulation between IC-FERST and PHREEQC to generate the dataset, since the goal is to use the machine learning surrogate to replace PHREEQC in the coupled simulation. The dataset used in the previous section was generated using Monte Carlo simulation and a uniform distribution with the same fixed range for all the input variables (see Section~\ref{sec:mlm}). In this section, we test the following sampling procedures to generate the inputs for the datasets:
\begin{enumerate}

    \item Using Monte Carlo to sample the inputs from a uniform distribution with the same fixed range (concentrations from 0.0 to 0.0015 mol/kgw). We will call this ``vanilla'' sampling. 
    \item Using Monte Carlo to sample the inputs from uniform distributions but using the concentration ranges from  Figure~\ref{fig:orig}. 
    \item Using Monte Carlo to sample the inputs from a uniform distribution with the same fixed range and randomly enforcing the concentration zero in 30\% of the sample inputs. We did this because looking at Figure~\ref{fig:orig}, we see that most of the time one or more cations have zero concentration during the simulation.   
    \item Using Monte Carlo to sample the inputs from uniform distributions but using the concentration ranges from  Figure~\ref{fig:orig}, and randomly enforcing the concentration zero in 30\% of the sample inputs.
    \item Using Monte Carlo to sample the inputs from a given covariance matrix. The assumption here is that we know the relation between the inputs and can provide the corresponding covariances.   
\end{enumerate}

\begin{figure}[!tb]
	\centering
	\includegraphics[width=1.0\linewidth]{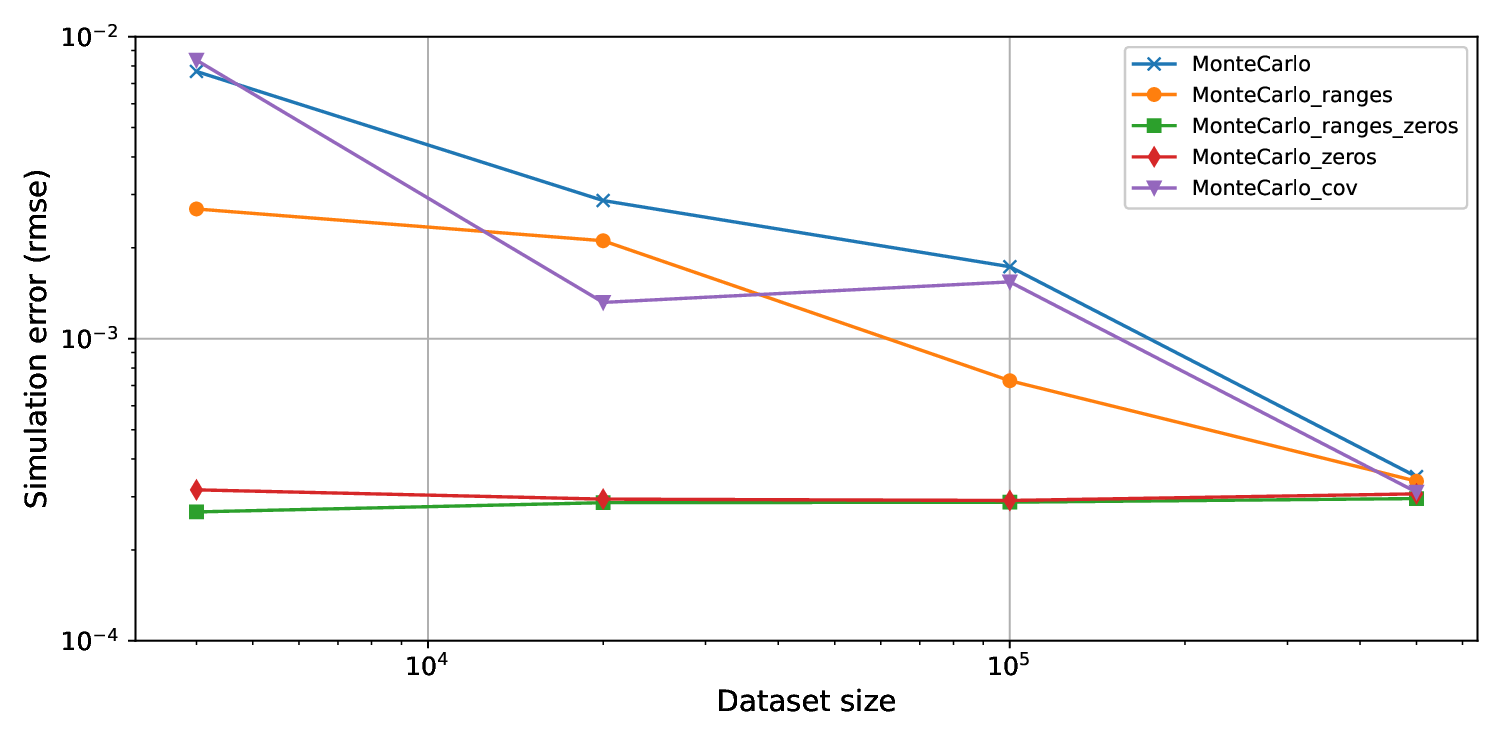}
	\caption{Comparison of different sampling procedures and dataset sizes in terms of rollout/simulation error.}
	\label{fig:ds}
\end{figure} 

For each one of these sampling procedures, we train the XGBoost with residual connection using different dataset sizes, ranging from 4,000 to 500,000 instances. Figure~\ref{fig:ds} shows the rollout/simulation error for all these cases.  We notice that as more information/physics is added to the training samples (input space) better the result. The ``vanilla'' sampling produces the worst results, and the process of enforcing zeros is the best one. We can also see that if we do not have enough information about the input space, adding more instances to the dataset can definitely improve the results. On the other hand, for the cases where we enforce the zeros a much smaller dataset can be used for training. The case using the covariance matrix does not produce great results, this indicates that the manifold (in the input space) generated by the physically plausible inputs cannot be described only using the covariances.    

\begin{figure}[!tb]
	\centering
	\includegraphics[width=0.9\linewidth]{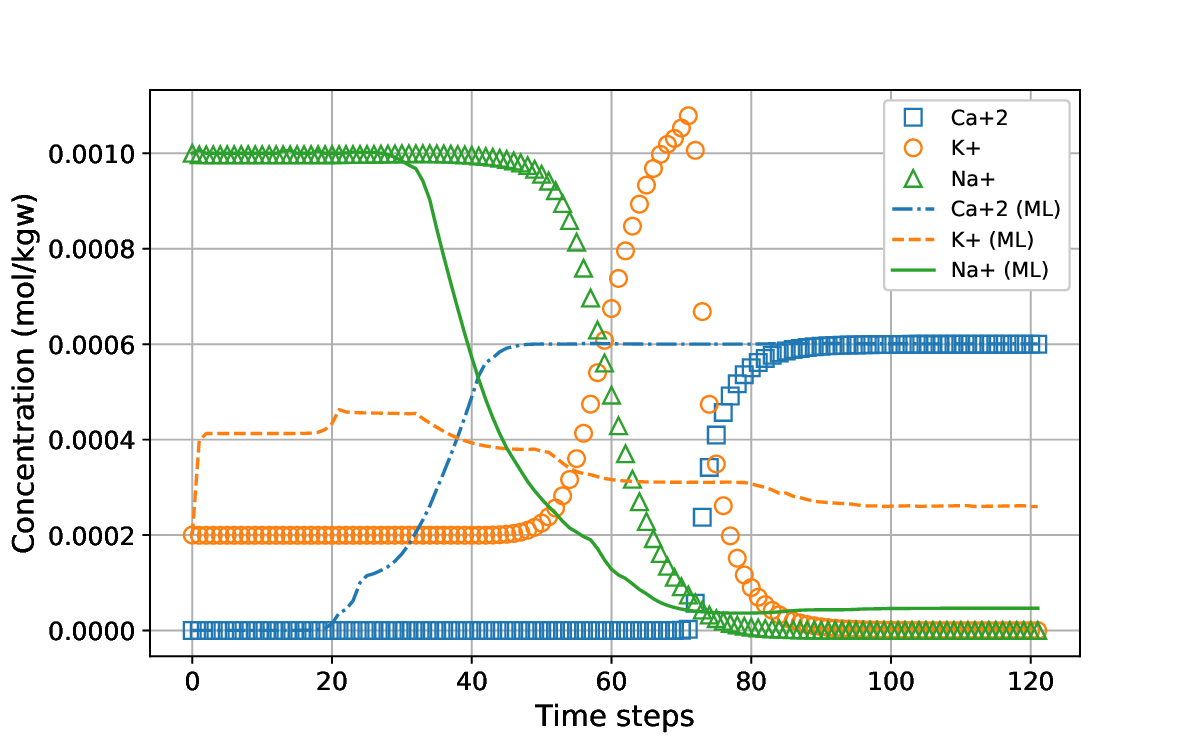}
	\caption{Simulation results comparing the coupling between PHREEQC (circles) and the machine learning surrogate (lines) with IC-FERST. We train the machine learning using the dataset enforcing the zeros.}
	\label{fig:pred2}
\end{figure}

Figure~\ref{fig:pred2} shows the simulation results, but now training the XGBoost with residual connections with the dataset enforcing the zeros (one of the lowest errors in Figure~\ref{fig:ds}). We train the machine learning model using 100,000 instances in the dataset. The results from Figure~\ref{fig:pred2} are better (in terms of squared error) than Figure~\ref{fig:pred1}; however, the machine learning surrogate is still not able to replace the geochemical simulator. For the remainder of the paper, we are going to use the dataset enforcing the zeros.

\subsubsection{Physical constraints}

\begin{figure}[!tb]
	\centering
	\includegraphics[width=\textwidth]{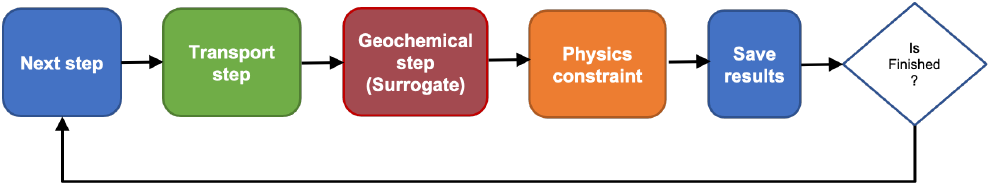}
	\caption{Coupling between the flow and transport simulator (IC-FERST) and the machine learning surrogate. We add a post-processing step to guarantee mass/charge balance.}
	\label{fig:fcfinal}
\end{figure}

In the previous section, we observed that the more information/physics we add to the training samples, the lower the simulation/rollout error will be. Based on this, we decided to include more prior knowledge to the coupling procedure. The idea is to enhance the coupling between IC-FERST and the machine learning surrogate with the physics of the problem. Below, we list the modifications to the coupling procedure:

\begin{enumerate}
    \item When the concentrations of the cations in one grid cell is equal to the concentrations in the inflow of that same cell, we do not call the surrogate. We assume that the output (final solution) has the same concentrations. No geochemical reaction will occur in this case because the solution is in equilibrium. 
    \item Every $n-1$ time steps (here $n=10$), we use the geochemical simulator instead of the machine learning surrogate in a new time step. This procedure enable us to correct any small deviation in the solution caused by the surrogate. In terms of computational cost, we are still replacing PHREEQC in 90\% of the callings. 
    \item We enforce the charge balance - the charge on both sides of the Eq.~\ref{eq:cex} must be equal because the reaction is stoichiometrically balanced - in the output of the machine learning surrogate, as a post-processing step (see Figure~\ref{fig:fcfinal}). We linearly rescale the concentrations in the output to match the same charge in the input solution. 
\end{enumerate}

\begin{figure}[!tb]
	\centering
	\includegraphics[width=0.9\linewidth]{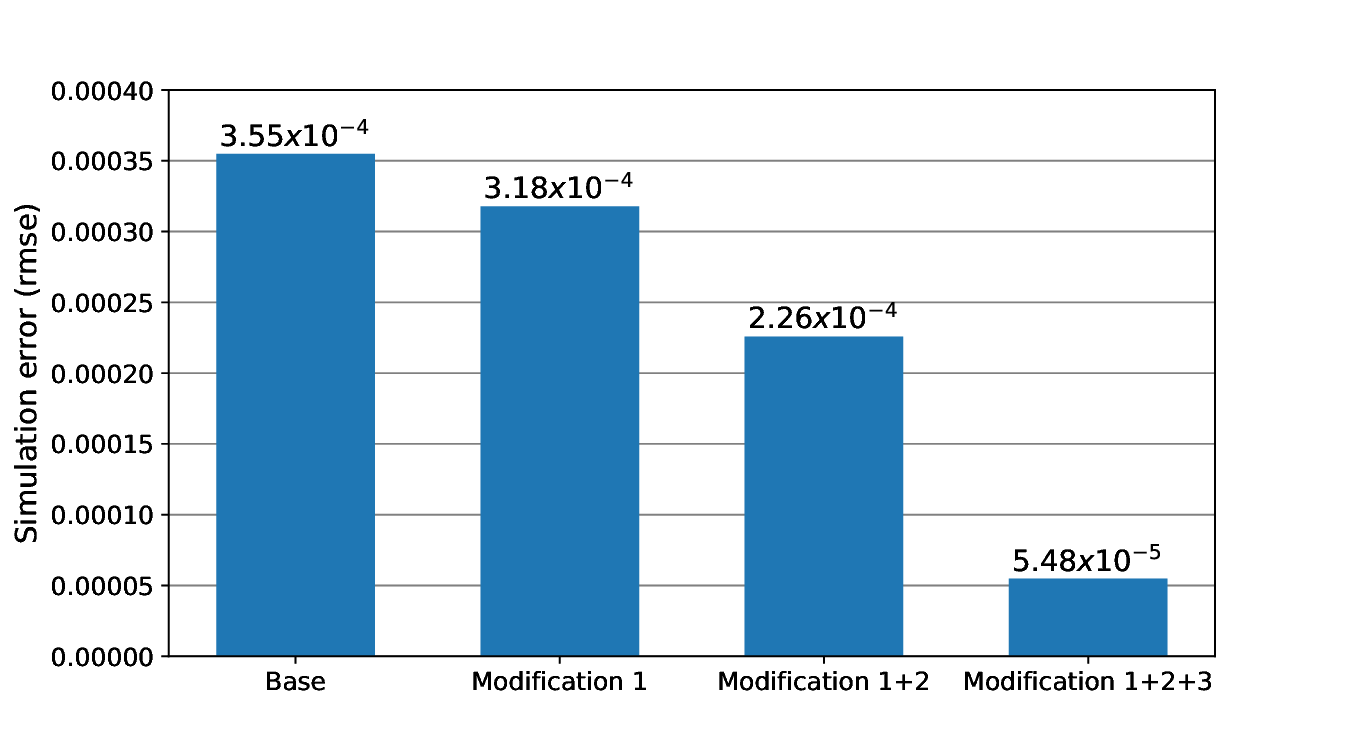}
	\caption{Adding the prior/physics knowledge into the coupling procedure. If we apply the modifications in a different order, the charge balance still generates the greatest decrease in error followed by the periodic geochemical simulator call. We still need the three modifications to achieve the lowest level of mismatch.}
	\label{fig:rmse_physics}
\end{figure}

\begin{figure}[!tb]
	\centering
	\begin{subfigure}[t]{0.7\columnwidth}
		\centering
		\includegraphics[width=\columnwidth]{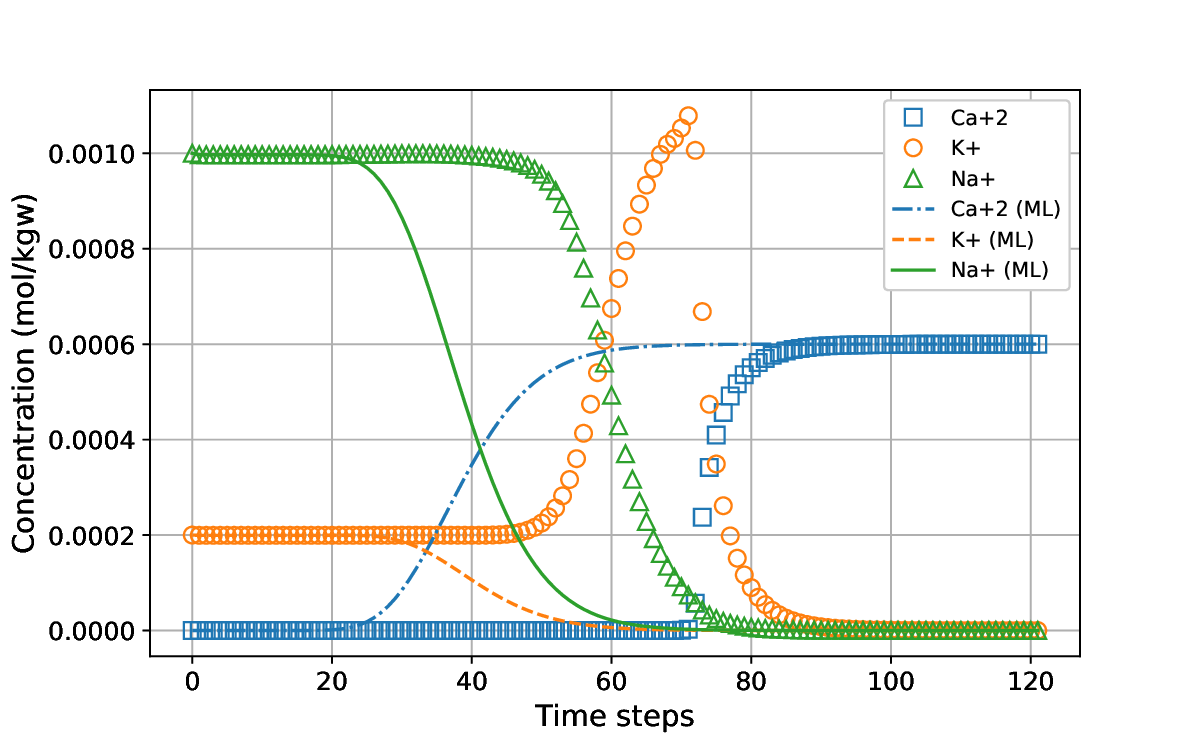}
		\caption{Adding modification 1}
		\label{fig:noth}
	\end{subfigure}
	\begin{subfigure}[t]{0.7\columnwidth}
		\centering
		\includegraphics[width=\columnwidth]{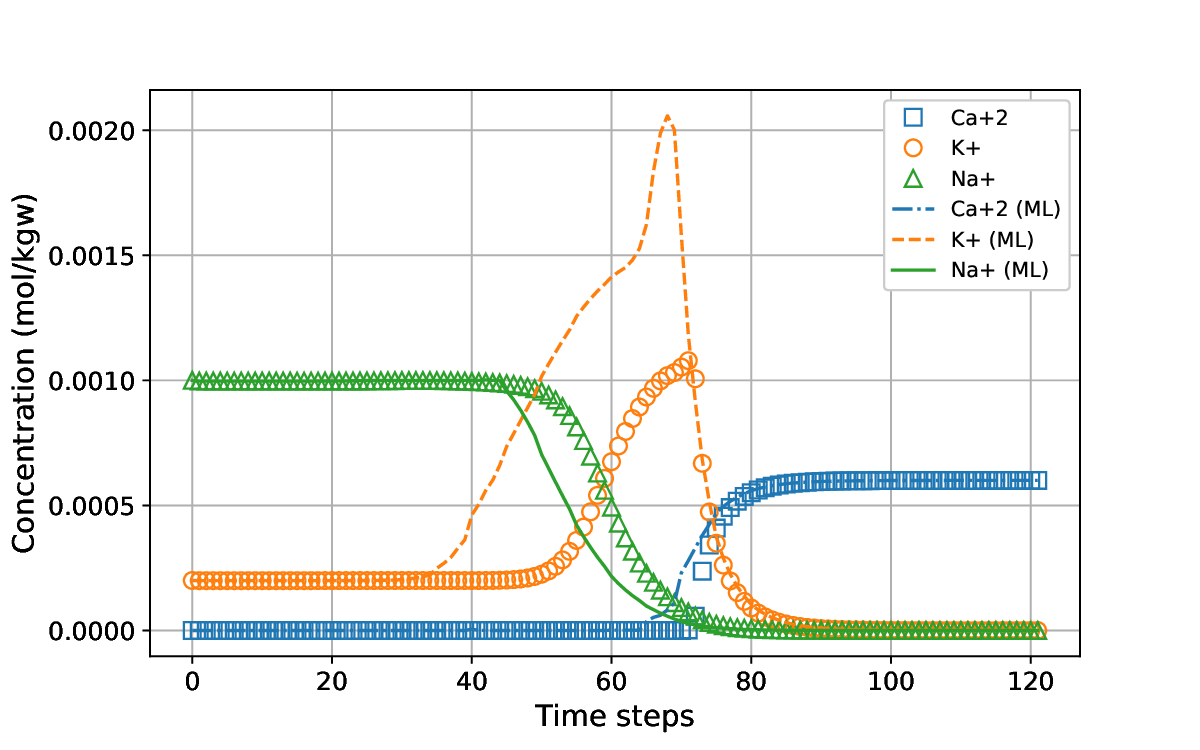}
		\caption{Adding modifications 1 and 2.}
		\label{fig:noth_plus_up}
	\end{subfigure}
	\caption{Simulation results comparing the coupling between PHREEQC (circles) and the machine learning surrogate (lines) with IC-FERST.}
	\label{fig:noth_up}
\end{figure}

We show in Figure~\ref{fig:rmse_physics}, the rollout/simulation error of the coupling between IC-FERST and the machine learning surrogate adding each one of these modifications. For the first modification, we observe a small decrease in the error. Figure~\ref{fig:noth} shows the simulation results comparing the coupling between the geochemical simulator and the machine learning surrogate with IC-FERST, considering that we do not call the surrogate (or the geochemical simulator) when the inflow cation concentrations are equal to the concentrations already in the grid cell. We can notice in this case that the cation concentrations at the beginning of the simulation match the ground truth. Because we inject on the left of the domain and produce on the right (see Figure~ \ref{fig:domain}), in the first time steps the concentration in the output will be equal to the initial one, which means no geochemical reactions need to be performed. For the second modification, we see a higher decrease in the error in Figure~\ref{fig:rmse_physics}. We show in Figure~\ref{fig:noth_plus_up} the simulation results considering the two modifications. These results are much closer to the ground truth, although not yet being able to replace it.            

\begin{figure}[!tb]
	\centering
	\includegraphics[width=0.9\linewidth]{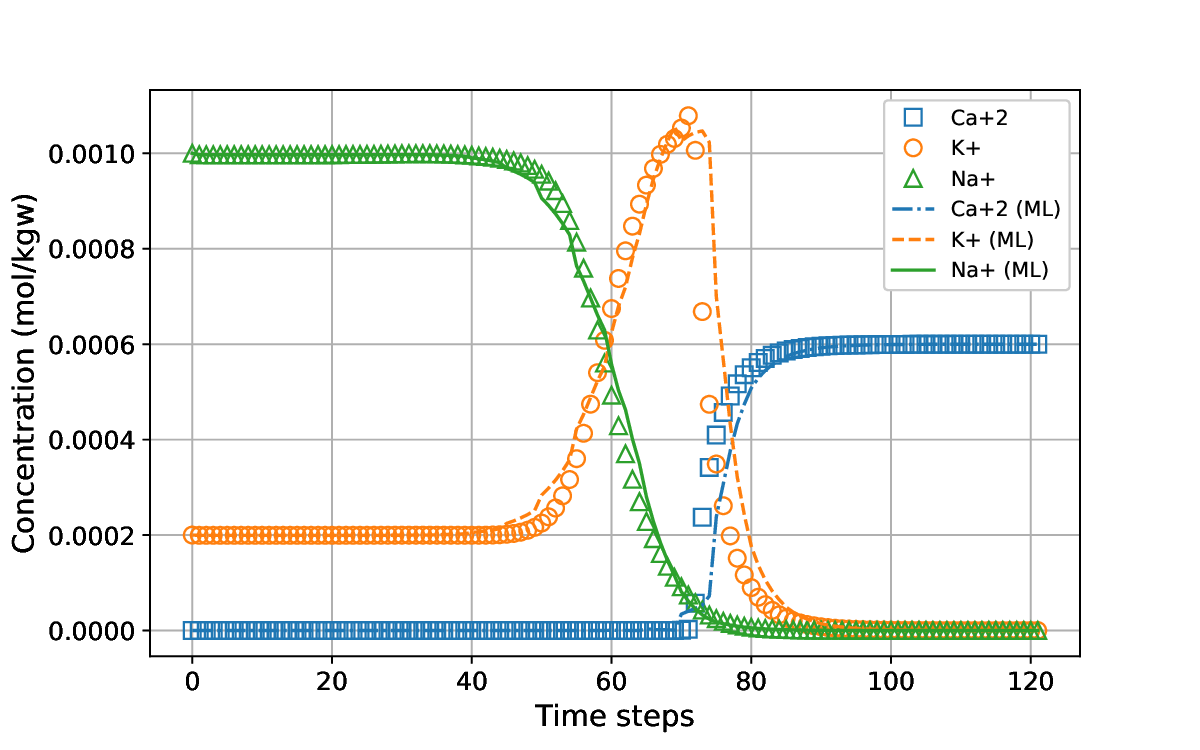}
	\caption{Simulation results comparing the coupling between PHREEQC (circles) and the machine learning surrogate (lines) with IC-FERST. Adding the prior knowledge/physics in modifications 1, 2 and 3.}
	\label{fig:predfinal}
\end{figure}

We also notice from Figure~\ref{fig:noth_plus_up} that the amount of \ce{K+} is much higher when we use the machine learning surrogate rather than the geochemical simulator. It indicates that the surrogate is generating mass/charge, in other words,  creating more \ce{K+} than what was injected or present in the domain. Considering this, we add a mass balance (here in the form of charge balance) as a post-processing step of the machine learning surrogate, as shown in Figure~\ref{fig:fcfinal}. Because the reaction in Eq.~\ref{eq:cex} is stoichiometrically balanced, the charge on both sides of the equation must be equal. Therefore, we enforce the charge balance in the output of the machine learning model. We just rescale the output concentrations to conserve the charge balance in the solution (so that the charge of the output solution matches that of the input solution). We see from Figure~\ref{fig:rmse_physics} that the difference in error due to the charge balance is the greatest one, the error is almost one order of magnitude lower than the other ones. Figure~\ref{fig:predfinal} shows the simulation results after adding the three modifications. We can see a good match between the coupling using the machine learning surrogate and the geochemical simulator. If we remove any one of the three modifications, we do not generate a good match. It is worth noticing that the surrogate (XGBoost with residual connection) runs 10 times faster than the geochemical simulator and that we replace the latter in 90\% of the callings.     

\section{Discussion}\label{sec:disc}


We showed that the use of machine learning surrogates for modelling reactive transport cannot be seen as only fitting a good regression model. The rollout has a huge effect on the final result, and the physics of the problem needs to be taken into account. For the geochemical calculations in the cation exchange problem, we could achieve a speed-up of at least one order of magnitude which can enable the coupling for many reactive transport problems, since the reaction cost is a deterrent. It is worth mentioning that this speed-up cannot be generalised for more complex geochemical calculation, other investigations should be done depending on the problem being solved. 
We tested the proposed approach in a cation exchange problem (sorption equilibrium reaction) as a proof-of-concept to thoroughly analyze the limitations and benefits of employing a machine learning surrogate. 

We demonstrated that ``classical'' machine learning methods still outperform recently proposed deep learning architectures for tabular data, and that XGBoost produced the more accurate results. Also, adding a residual/skip connection improves the prediction accuracy. Although the machine learning surrogate produced low errors in the one-shot prediction ($R^2=0.993$ in Figure~\ref{fig:pred}), when considering the coupling procedure and multiple time steps (rollout), the simulation using the surrogate fails (if nothing else is done). It seems that even small deviations in the surrogate outputs can be amplified during the rollout. We showed that the sampling procedure used to generate the dataset has a great impact on the simulation/rollout error. As much prior knowledge (physics) we add to the dataset better will be the results. Furthermore, incorporating this prior information into the training set can allow the use of much smaller dataset sizes than would be needed. On the other hand, if no prior knowledge is available including more data can reduce the simulation/rollout error.      

Even using a more appropriate dataset and the best machine learning model in a simple sorption equilibrium reaction, the surrogate was not able to replace the geochemical simulator for the rollout (without the physical constraints). The simulation error was still high. A crucial step for devising these surrogates is to incorporate the knowledge/physics into the modelling procedure. For the cation exchange problem, we demonstrated that adding three small changes into the simulation workflow, being the mass/charge balance the most important one (Figure~\ref{fig:fcfinal}), can make the machine learning surrogate able to replace the geochemical simulator. It is worth mentioning, that we also tested a physics-informed approach using a multilayer perceptron. However, the results were not as good as the model which used the physical constraints.         

\section{Conclusions}\label{sec:conc}

We evaluated machine learning models as replacements for the reaction module in a reactive transport simulation. The proposed approach is tested in a well-documented cation exchange problem chosen as a proof of concept to thoroughly analyze the limitations and benefits of employing a machine learning surrogate. We compared several machine learning models, with XGBoost with a residual/skip connection showing the best results in terms of prediction accuracy and being ten-fold faster than the geochemical simulator. However, when considering the rollout prediction over multiple time steps, the machine learning surrogate alone failed to match the accuracy of the geochemical simulator. To address this, we introduced simple non-intrusive modifications to improve the machine learning surrogate's performance in the rollout prediction: (i) not calling the surrogate when concentrations are in equilibrium, (ii) periodic validation against the geochemical simulator, and (iii) enforcing charge balance in the output. These modifications led to a significant reduction in the rollout/simulation error. Furthermore, the choice of dataset generation played a crucial role, highlighting the importance of incorporating prior knowledge or physics into the dataset sampling procedure. 
This work demonstrates that while machine learning surrogates can provide fast and accurate one-shot predictions, incorporating physics-based constraints and appropriate dataset generation strategies is essential for achieving reliable results in rollout predictions. By incorporating these modifications, we could replace the geochemical simulator for the cation exchange problem, while accelerating the geochemical calculations by at least one order of magnitude.  

\section*{Statements and Declarations}

\subsection*{Funding}
We would like to acknowledge financial support from Petrobras for the first author. 

\subsection*{Competing Interests}

The authors have no relevant financial or non-financial interests to disclose

\subsection*{Data Availability Statement}

The source code, data, and hardware configuration used in this work are available on Zenodo via \url{https://doi.org/10.5281/zenodo.14843955} \citep{softwarereact:2025}.



\bibliography{references}
\bibliographystyle{abbrvnat} 
~
\end{document}